\documentclass[pra,showpacs,twocolumn]{revtex4-1}%
\usepackage{amsfonts}
\usepackage{amsmath}
\usepackage{amssymb,amscd}
\usepackage{psfrag}
\usepackage{graphicx}
\usepackage{braket}
\usepackage[dvips]{epsfig}
\usepackage{epsfig}
\usepackage{subfigure}
\usepackage{color}
\usepackage{tabulary}
\usepackage{booktabs}
\usepackage{amssymb}%
\providecommand{\U}[1]{\protect\rule{.1in}{.1in}}
\begin{document}

\title{ Effects of losses on the sensitivity of an actively correlated Mach-Zehnder interferometer}
\author{Qiang Wang$^{1}$}
\author{Gao-Feng Jiao$^{1}$}
\author{Zhifei Yu$^{1}$}
\author{L. Q. Chen$^{1,3}$}
\email{lqchen@phy.ecnu.edu.cn}
\author{Weiping Zhang$^{2,3,4}$}
\author{Chun-Hua Yuan$^{1}$}
\email{chyuan@phy.ecnu.edu.cn}
\address{$^1$State Key Laboratory of Precision Spectroscopy, Quantum Institute for Light and Atoms, Department of Physics, East China Normal University, Shanghai 200062, China}

\address{$^2$School of Physics and Astronomy, and Tsung-Dao Lee Institute, Shanghai Jiao Tong University, Shanghai 200240, China}

\address{$^{3}$Shanghai Research Center for Quantum Sciences, Shanghai 201315, China}

\address{$^{4}$Collaborative Innovation Center of Extreme Optics, Shanxi
University, Taiyuan, Shanxi 030006, China}

\begin{abstract}
We theoretically studied the quantum Cram\'{e}r-Rao bound of an actively
correlated Mach-Zehnder interferometer (ACMZI), where the quantum Fisher
information obtained by the phase-averaging method can give the proper
phase-sensing limit without any external phase reference. We numerically
calculate the phase sensitivities with the method of homodyne detection and
intensity detection in the presence of losses. Under lossless and very low
loss conditions, the ACMZI is operated in a balanced case to beat the standard
quantum limit (SQL). As the loss increases, the reduction in sensitivity
increases. However within a certain range, we can adjust the gain parameters
of the beam recombination process to reduce the reduction in sensitivity and
realize the sensitivity can continue to beat the SQL in an unbalanced situation.
Our scheme provides an optimization method of phase estimation in the presence
of losses.
\end{abstract}
\date{\today }
\maketitle

\section{Introduction}

The fundamental setup of a metrology device is a Mach-Zehnder interferometer
(MZI), which has been used in many fields such as phase estimation and
gravitational wave detection \cite{MZ1,MZ2}. Sensitivity is one of the most
important indexes of an interferometer. However in a generic interferometric
measurement using an MZI and classical light sources the precision of
estimating the relative phase delay $\phi$ inside the interferometer is
bounded by $\Delta^{2}\phi\geq1/n^{2}$, where $n$ is the average photon number
inside interferometer. This bound is referred to as the shot-noise limit or
standard quantum limit (SQL) \cite{MZ3}.

In order to improve the sensitivity, researchers have proposed a number of
schemes. The first way is to use quantum states to beat SQL such as squeezed
states \cite{MZ3}, N00N states \cite{MZ4}, twin Fock states \cite{MZ5}, and
two-mode squeezed states \cite{MZ6}. The other way of beating SQL is to use
active elements in an interferometer. SU(1,1) interferometer was proposed to
one of them by Yurke in 1986 \cite{Yurke86}. For SU(1,1) interferometer, the
beam splitters in the MZI are replaced by nonlinear beam splitter such as an
optical parametric amplifier (PA) or a four-wave mixers, which are
mathematically characterized by the group SU(1,1). Because the sensitivities
of these interferometers can achieve Heisenberg limit, this type of
interferometers have received extensive attention both experimentally
\cite{su1,su2,su3,su4,su5,su6,su7,su8,su9,su10} and theoretically
\cite{su11,su12,su13,su14,su15,su16,su17,su18,su19,su20,su21,Caves20}.

Combination of the quantum state input and active elements in interferometers,
a new variant of MZI, actively correlated Mach-Zehnder interferometer (ACMZI)
recently was proposed \cite{MZ8}. Compared to the traditional MZI, in addition
to a coherent state in one input port, one mode of a two-mode squeezed-vacuum
state is in the other input port and the output is detected with the active
elements. That is the final interference output of the MZI is detected with
the method of active correlation output readout.

In this paper, we theoretically derive the quantum Cram\'{e}r-Rao bound (QCRB)
of ACMZI according to the quantum Fisher information (QFI) obtained by the
phase-averaging method, which can give the proper phase-sensing limit without
any external phase reference. To approach the QCRB, we calculate the phase
sensitivities with the method of homodyne detection (HD) and intensity
detection (ID) in the presence of losses. Under lossless and very low loss
conditions, the ACMZI is operated in a balanced case to beat the SQL. As the
loss increases, the reduction in sensitivity increases. We can adjust the gain
parameters of the beam recombination process to reduce the reduction in
sensitivity and realize the sensitivity can continue to beat the SQL within a
certain range. The corresponding optimization conditions are given.

Our paper is organized as follows. In Sec. II, the QCRB of ACMZI is derived
according to the phase-averaging method. In Sec. III, the phase sensitivity is
studied with the method of HD and ID in the presence of losses. Due to losses
the sensitivity is reduced in a balanced case, and we describe that the
sensitivity can continue to beat the SQL in an unbalanced situation by
optimizing the gain ratio. Finally, our results are summarized.

\begin{figure}[tb]
\centering{\includegraphics[scale=0.35,angle=0]{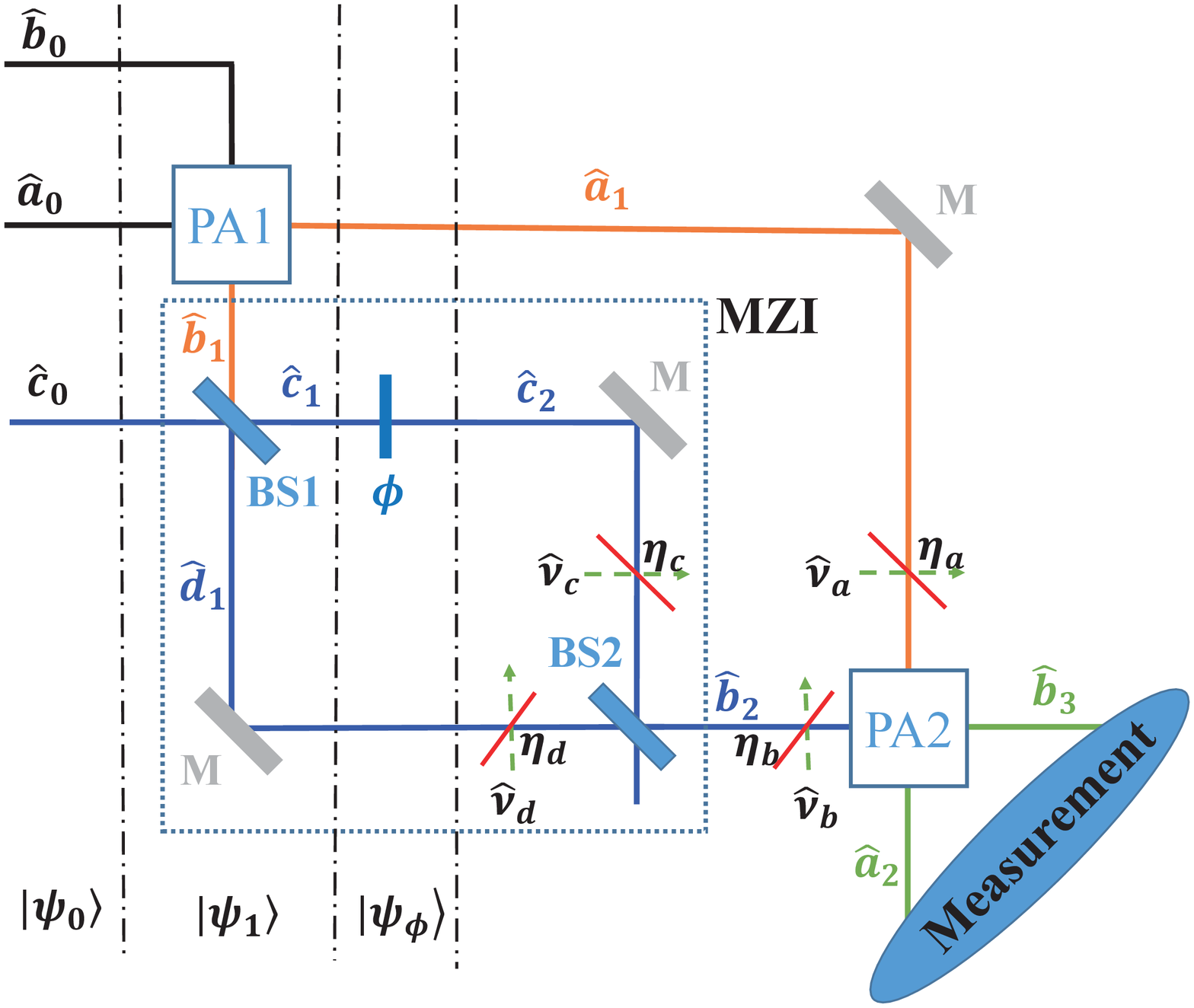}} \caption{ Scheme of
an actively correlated Mach-Zehnder interferometer. One of the two-mode
quantum states generated by PA1 is input to the MZI in the dashed box, and the
other mode of the two-mode squeezed state are combined by with PA2 to realize
enhanced readout. By adding the fictitious beam splitters, the arms of the MZI
has different transmission rates, where $\eta_{a}$, $\eta_{b}$ are the
external transmission rates of the MZI, and $\eta_{c}$, $\eta_{d}$ are the
internal transmission rates of the MZI. $\hat{a}_{i}$, $\hat{b}_{i}$, $\hat
{c}_{i}$ and $\hat{d}_{i}$ ($i=0,1,2,3$) denote light beams in the different
processes. $\mathrm{M}$: mirrors; $\mathrm{BS}$: linear beam splitters;
$\mathrm{PA}$: parametric amplifier. }%
\label{fig1}%
\end{figure}

\section{QCRB of ACMZI}

\subsection{Input-output relation of ACMZI}

In this section, we review the input-output relation of ACMZI. As is shown in
Fig.~\ref{fig1} \cite{MZ8}, one input port of MZI is injected with a strong
pump field, and another input port is sent with one mode of a two-mode
entanglement state generated by PA1. The unknown phase shift is embedded in
one arm of MZI as the estimator. One of the output fields of the MZI and the
other mode of the two-mode squeezed state are combined by with PA2 to realize
enhanced readout.

The four modes in the ACMZI are described by the annihilation operators
$\hat{a}_{i}$, $\hat{b}_{i}$, $\hat{c}_{i\text{ }}$and $\hat{d}_{i}$ $\left(
i=0,1,2,3\right)  $. The phase shift is described as $U\left(  \phi\right)
=e^{i\phi\hat{n}_{c_{1}}}$. In the Heisenberg picture, the output ports
$\hat{a}_{2}$\ and $\hat{b}_{3}$ described by%
\begin{align}
\hat{a}_{2} &  =T_{1}\hat{a}_{0}+T_{2}\hat{b}_{0}^{\dagger}+T_{3}\hat{c}%
_{0}^{\dagger},\\
\hat{b}_{3} &  =M_{1}^{\ast}\hat{a}_{0}^{\dagger}+M_{2}^{\ast}\hat{b}%
_{0}+M_{3}^{\ast}\hat{c}_{0},
\end{align}
with
\begin{align}
T_{1} &  =G_{2}G_{1}+g_{1}g_{2}e^{i(\theta_{2}-\theta_{1})}(T-Re^{-i\phi
}),\nonumber\\
T_{2} &  =G_{2}g_{1}e^{i\theta_{1}}+G_{1}g_{2}e^{i\theta_{2}}(T-Re^{-i\phi
}),\nonumber\\
T_{3} &  =\sqrt{TR}g_{2}e^{i\theta_{2}}(1+e^{-i\phi}),\nonumber\\
M_{1}^{\ast} &  =G_{1}g_{2}e^{i\theta_{2}}+G_{2}g_{1}e^{i\theta_{1}%
}(T-Re^{i\phi}),\nonumber\\
M_{2}^{\ast} &  =g_{1}g_{2}e^{i(\theta_{2}-\theta_{1})}+G_{2}G_{1}%
(T-Re^{i\phi}),\nonumber\\
M_{3}^{\ast} &  =\sqrt{TR}G_{2}(1+e^{i\phi}),
\end{align}
where $R$ and $T$ are the reflectivity and transmissivity of the two BSs,
$G_{1}$ and $G_{2}$ are the gain factors of PA1 and PA2, for wave splitting
and recombination with $G_{i}^{2}-g_{i}^{2}=1$ $(i=1,2)$, $\theta_{1}$ and
$\theta_{2}$ describe the phase shifts of the PAs for wave splitting and
recombination, respectively.

\subsection{Phase averaged QFI}

In this section, the QFI and phase averaged QFI of ACMZI are given,
respectively. The QFI is the intrinsic information about the quantum state and
is not related to the actual measurement procedure. The phase estimation bound
can be dealt with the method of QFI \cite{QFI00,QFI01}.

We consider one input $\hat{c}_{0}$ is a coherent state $|\alpha\rangle$, and
the other two input $\hat{a}_{0}$ and $\hat{b}_{0}$ is a vacuum state. As
shown in Fig~\ref{fig1}, the probe state is a pure state $|\psi_{0}\rangle$,
and after injecting into the PA1 and BS1 the state transform into the
correlated probe state $|\psi_{1}\rangle$, and the probe state $|\psi
_{1}\rangle$ is modified as $|\psi_{\phi}\rangle$ after phase shifted. For the
symmetric logarithmic derivatives (SLD) operator, the QFI is $F=4(\langle
\partial_{\phi}\psi_{\phi}|\partial_{\phi}\psi_{\phi}\rangle-\left\vert
\langle\partial_{\phi}\psi_{\phi}|\psi_{\phi}\rangle\right\vert ^{2})$, where
$|\partial_{\phi}\psi\rangle=\partial|\psi\rangle/\partial_{\phi}$
\cite{QFI02,QFI03}. After the calculation, the QFI is given by%
\begin{align}
F  &  =N_{c}\left[  4T^{2}+4TR\left(  G_{1}^{2}+g_{1}^{2}\right)  \right]
\nonumber\\
&  +g_{1}^{2}\left[  4R^{2}G_{1}^{2}+4TR\right]  ,
\end{align}
where $N_{c}=\langle\alpha|\hat{n}_{c_{0}}|\alpha\rangle$. When $N_{c}\gg
G_{1}^{2}$, $F\approx N_{c}\left[  4T^{2}+4TR\left(  G_{1}^{2}+g_{1}%
^{2}\right)  \right]  $.

As pointed by Jarzyna et al. \cite{QFI2}, the QFI-only approach may have the
pitfall that the optimal POVM attaining the QCRB might contain huge amounts of
hidden resources. The remedy is to exclude any external resources that might
provide some phase information to the measuring device when implementing the
best POVM. Such a \textquotedblleft rule-out\textquotedblright\ protocol was
introduced by Jarzyna \cite{QFI2}, where the issue is resolved by introducing
the phase-averaging of the three-mode input state via a common phase shift.
The QFI obtained by the phase-averaging method can give the proper
phase-sensing limit without any external phase reference \cite{QFI1,QFI3}.

Let us expand the input state $|\psi_{0}\rangle\langle\psi_{0}|$ in the
photon-number basis
\begin{equation}
|\psi_{0}\rangle\langle\psi_{0}|=\sum\limits_{nm}c_{n}c_{m}|n\rangle\langle
m|_{C}\otimes|0\rangle\langle0|_{B}\otimes|0\rangle\langle0|_{A},
\end{equation}
where $|n\rangle$ is the photon number state. The reference framed between the
inputs and the measurement is removed by phase averaging the input state as
\cite{QFI1,QFI3}
\begin{align}
|\psi_{0}\rangle\langle\psi_{0}|_{ave}  &  =\int\frac{d\theta}{2\pi}V_{\theta
}^{A}V_{\theta}^{B}V_{\theta}^{C}|\psi_{0}\rangle\langle\psi_{0}|V_{\theta
}^{A\dag}V_{\theta}^{B\dag}V_{\theta}^{C\dag}\nonumber\\
&  =\sum\limits_{nm}\int\frac{d\theta}{2\pi}c_{n}c_{m}^{\ast}e^{i\theta\left(
n-m\right)  }|00n\rangle\langle00m|\nonumber\\
&  =\sum\limits_{n}P_{n}|n\rangle\langle n|_{C}\otimes|0\rangle\langle
0|_{B}\otimes|0\rangle\langle0|_{A}\nonumber\\
&  =\sum\limits_{n}P_{n}|00n\rangle\langle00n|,
\end{align}
where $V_{\theta}^{A}=e^{i\theta a^{\dag}a}$, $V_{\theta}^{B}=e^{i\theta
b^{\dag}b}$, $V_{\theta}^{C}=e^{i\theta c^{\dag}c}$, $P_{n}=c_{n}c_{m}^{\ast}%
$, and $\sum\limits_{n}P_{n}=1$. The output state is given by%
\begin{equation}
|\psi_{\phi}\rangle\langle\psi_{\phi}|_{ave}=T_{T,g}^{\phi}\left(  |\Psi
_{0}\rangle\langle\Psi_{0}|_{ave}\right)  T_{T,g}^{\dagger\phi}=\sum
\limits_{n}P_{n}|\psi_{n}\rangle\langle\psi_{n}|,
\end{equation}
where $|\psi_{n}\rangle=T_{T,g}^{\phi}|00n\rangle=e^{i\phi n_{c}}%
B_{T,g}|00n\rangle$, $B_{T,g}$ represents the combined action of PA1 and BS1.

Due to the convexity of the QFI \cite{QFI4}, we can obtain the total QFI of
the above density matrix, which is given by
\begin{align}
F_{ave}  &  =\sum\limits_{n}P_{n}F\left(  |\psi_{n}\rangle\right) \nonumber\\
&  =4R^{2}G_{1}^{2}g_{1}^{2}+4TR\left[  N_{C}G_{1}^{2}+g_{1}^{2}\left(
N_{C}+1\right)  \right] \nonumber\\
&  =4N_{c}TR\left(  G_{1}^{2}+g_{1}^{2}\right)  +g_{1}^{2}\left[  4R^{2}%
G_{1}^{2}+4TR\right]  .
\end{align}
Because of $N_{c}\gg G_{1}^{2}$, we rewrite the above equation as
$F_{ave}\approx4N_{c}TR\left(  G_{1}^{2}+g_{1}^{2}\right)  $. Comparing $F$
with $F_{ave}$, we can simply find that the fluctuation terms of the coherent
light $\left(  4T^{2}N_{c}\right)  $ are eliminated. The $F_{ave}$ value of
phase averaging is more accurate, then we use $F_{ave}$ as our bound. When
$T=R=0.5$, the value of $F_{ave}$ is the maximum value:%
\begin{equation}
\max F_{ave}=N_{c}\left(  G_{1}^{2}+g_{1}^{2}\right)  .
\end{equation}
The QCRB states that whatever the measurement chosen, the following bound on
the estimation uncertainty holds
\begin{equation}
\Delta^{2}\phi\geq\frac{1}{F_{ave}}.
\end{equation}

\begin{figure}[tb]
\centering{\includegraphics[scale=0.33,angle=0]{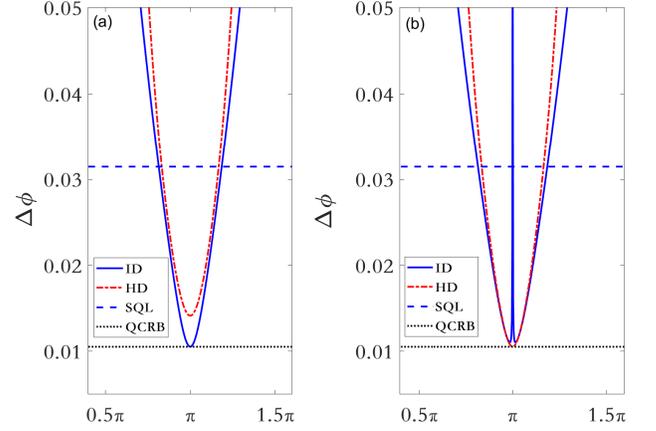}}\caption{ Phase
sensitivity as a function of $\phi$ (a) balanced case and (b) unbalanced case
with $G_{2}^{2}=20$. Parameters: $N_{C}=1000$, $G_{1}^{2}=5$, and $T=R=0.5$.}%
\label{fig2}%
\end{figure}

\section{Phase Sensitivities}

In this section, we study the ultimate precision limit of ACMZI in noisy
metrology and the effects of losses on phase sensitivities $\Delta\phi$ and
optimal gain ratio $(G_{2}/G_{1})_{opt}$. The QFI represents the maximum
amount of information that can be extracted from quantum experiments. However,
saturating the limit obtained by QFI is an important issue. Here, we consider
two detection methods: ID and HD. Through an error propagation analysis the
sensitivity is given by \cite{Dowling08}
\begin{equation}
\Delta^{2}\phi=\frac{\langle\Delta^{2}\hat{O}\rangle}{\left\vert
\partial\langle\hat{O}\rangle/\partial\phi\right\vert ^{2}}, \label{error}%
\end{equation}
where $\langle\Delta^{2}\hat{O}\rangle=\langle\hat{O}^{2}\rangle-\langle
\hat{O}\rangle^{2}$, $\langle\Delta^{2}\hat{O}\rangle$ denotes the noise of
observable $\hat{O}$, and $|\partial\langle\hat{O}\rangle/\partial\phi|$ is
the slope with respect to the corresponding phase shift.

\begin{figure}[tb]
\centering{\includegraphics[scale=0.45,angle=0]{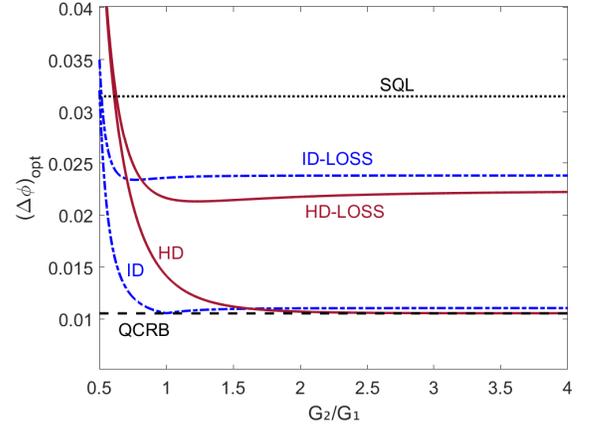}} \caption{ Optimum
phase sensitivity as a function of $G_{2}/G_{1}$. Parameters: $G_{1}^{2}=5$,
$N_{c}=1000$ and $T=R=0.5$. }%
\label{fig3}%
\end{figure}

By adding the fictitious beam splitters, the arms of the MZI has different
transmission rates, where $\eta_{a}$, $\eta_{b}$ are the external transmission
rates of the MZI, and $\eta_{c}$, $\eta_{d}$ are the internal transmission
rates of the MZI, as shown in Fig.~\ref{fig1}. Then the transforms of the
fields is given by%
\begin{align}
\hat{a}_{1}^{\prime} &  =\sqrt{\eta_{a}}\hat{a}_{1}+\sqrt{1-\eta_{a}}\hat
{v}_{a}\text{, }\hat{b}_{2}^{\prime}=\sqrt{\eta_{b}}\hat{b}_{2}+\sqrt
{1-\eta_{b}}\hat{v}_{b}\nonumber\\
\hat{c}_{2}^{\prime} &  =\sqrt{\eta_{c}}\hat{c}_{2}+\sqrt{1-\eta_{d}}\hat
{v}_{c}\text{, }\hat{d}_{1}^{\prime}=\sqrt{\eta_{d}}\hat{d}_{1}+\sqrt
{1-\eta_{d}}\hat{v}_{d},
\end{align}
where the $\hat{v}_{i}$ $\left(  i=a,b,c,d\right)  $ represent the vacuum.
Therefore, the input-output relation is written as%
\begin{align}
\hat{a}_{2l} &  =T_{1L}\hat{a}_{0}+T_{2L}^{\dagger}\hat{b}_{0}^{\dagger
}+T_{3L}\hat{c}_{0}^{\dagger}+T_{4}\hat{v}_{a}+T_{5}\hat{v}_{b}^{\dagger
}+T_{6}\hat{v}_{c}^{\dagger}\nonumber\\
&  +T_{7}\hat{v}_{d}^{\dagger},\\
\hat{b}_{3l} &  =M_{1L}^{\ast}\hat{a}_{0}{}^{\dagger}+M_{2L}^{\ast}\hat{b}%
_{0}+M_{3L}^{\ast}\hat{c}_{0}+M_{4}^{\ast}\hat{v}_{a}^{\dagger}+M_{5}^{\ast
}\hat{v}_{b}\nonumber\\
&  +M_{6}^{\ast}\hat{v}_{c}+M_{7}^{\ast}\hat{v}_{d},
\end{align}
where
\begin{align}
T_{1l} &  =G_{1}G_{2}\sqrt{\eta_{a}}+g_{1}g_{2}e^{i\left(  \theta_{2}%
-\theta_{1}\right)  }\sqrt{\eta_{b}}\left(  T\sqrt{\eta_{d}}-R\sqrt{\eta_{c}%
}e^{-i\phi}\right)  ,\nonumber\\
T_{2l} &  =G_{2}g_{1}\sqrt{\eta_{a}}e^{i\theta_{1}}+G_{1}g_{2}e^{i\theta_{2}%
}\sqrt{\eta_{b}}\left(  T\sqrt{\eta_{d}}-R\sqrt{\eta_{c}}e^{-i\phi}\right)
,\nonumber\\
T_{3l} &  =\sqrt{TR}g_{2}e^{i\theta_{2}}\left(  \sqrt{\eta_{b}\eta_{d}}%
+\sqrt{\eta_{b}\eta_{c}}e^{-i\phi}\right)  ,\nonumber\\
T_{4} &  =G_{2}\sqrt{1-\eta_{a}}\text{, }T_{6}=\sqrt{R}g_{2}\sqrt{\eta
_{b}\left(  1-\eta_{c}\right)  }e^{i\theta_{2}},\nonumber\\
T_{5} &  =g_{2}\sqrt{1-\eta_{b}}e^{i\theta_{2}}\text{, }T_{7}=\sqrt{T}%
g_{2}\sqrt{\eta_{b}\left(  1-\eta_{d}\right)  }e^{i\theta_{2}},\nonumber\\
M_{1l}^{\ast} &  =G_{1}g_{2}\sqrt{\eta_{a}}e^{i\theta_{2}}+G_{2}%
g_{1}e^{i\theta_{1}}\sqrt{\eta_{b}}\left(  T\sqrt{\eta_{d}}-R\sqrt{\eta_{c}%
}e^{i\phi}\right)  ,\nonumber\\
M_{2l}^{\ast} &  =g_{1}g_{2}\sqrt{\eta_{a}}e^{i\left(  \theta_{2}-\theta
_{1}\right)  }+G_{2}G_{1}\sqrt{\eta_{b}}\left(  T\sqrt{\eta_{d}}-R\sqrt
{\eta_{c}}e^{i\phi}\right)  ,\nonumber\\
M_{3l}^{\ast} &  =G_{2}\sqrt{TR}\sqrt{\eta_{b}}(\sqrt{\eta_{c}}e^{i\phi}%
+\sqrt{\eta_{d}}),\nonumber\\
M_{4}^{\ast} &  =g_{2}\sqrt{1-\eta_{a}}e^{i\theta_{2}}\text{, }M_{6}^{\ast
}=\sqrt{R}G_{2}\sqrt{\eta_{b}\left(  1-\eta_{c}\right)  },\nonumber\\
M_{5}^{\ast} &  =G_{2}\sqrt{1-\eta_{b}}\text{, }M_{7}^{\ast}=\sqrt{T}%
G_{2}\sqrt{\eta_{b}\left(  1-\eta_{d}\right)  }.
\end{align}

\subsection{Homodyne detection}

For the HD, the measurement operator is $\hat{Y}=i(\hat{b}_{3}^{\dagger}%
-\hat{b}_{3})$. Under the condition of $\theta_{1}=0$ and $\theta_{2}=\pi$,
the slope and variance are given by%
\begin{equation}
\frac{\partial\langle\hat{Y}\rangle}{\partial\phi}=2\sqrt{N_{C}}\sqrt{TR}%
G_{2}\sqrt{\eta_{b}\eta_{c}}\cos\phi, \label{slope}%
\end{equation}
and
\begin{align}
\langle\Delta^{2}\hat{Y}\rangle &  =2\eta_{a}G_{1}^{2}g_{2}^{2}+2T^{2}%
G_{2}^{2}g_{1}^{2}\eta_{b}\eta_{d}+2R^{2}G_{2}^{2}g_{1}^{2}\eta_{b}\eta
_{c}\nonumber\\
&  +4\left(  G_{1}\sqrt{\eta_{a}}g_{2}-TG_{2}g_{1}\sqrt{\eta_{b}}\sqrt
{\eta_{d}}\right) \nonumber\\
&  \times\left(  RG_{2}g_{1}\sqrt{\eta_{b}}\sqrt{\eta_{c}}\right)  \cos
\phi+2g_{2}^{2}\left(  1-\eta_{a}\right) \nonumber\\
&  -4TG_{1}G_{2}g_{1}g_{2}\sqrt{\eta_{b}}\sqrt{\eta_{d}}\sqrt{\eta_{a}}+1.
\label{variance}%
\end{align}
From Eq. (\ref{error}), we can obtain the phase sensitivity $\Delta\phi^{HD}$
in the presence of losses. From the phase sensitivity $\Delta\phi^{HD}$, the
optimal gain $G_{2}$ should satisfy the following condition%
\begin{equation}
\left(  \frac{g_{2}}{G_{2}}\right)  _{opt}=\frac{G_{1}g_{1}\sqrt{\eta_{a}%
}\sqrt{\eta_{b}}\left(  T\sqrt{\eta_{d}}+R\sqrt{\eta_{c}}\right)  }{\eta
_{a}G_{1}^{2}+\left(  1-\eta_{a}\right)  -1/2}.
\end{equation}
As the loss increases, the sensitivity reduction increases. However, we adjust
the gain ratio $G_{2}/G_{1}$ to reduce the reduction in sensitivity. Due to
the SQL is independent of $G_{2}$, we adjust the gain parameter $G_{2}$\ of
the beam recombination process to reduce the reduction in sensitivity and
realize that the sensitivity can continue to beat SQL within a certain range.

\begin{figure}[tb]
\centering{\includegraphics[scale=0.65,angle=0]{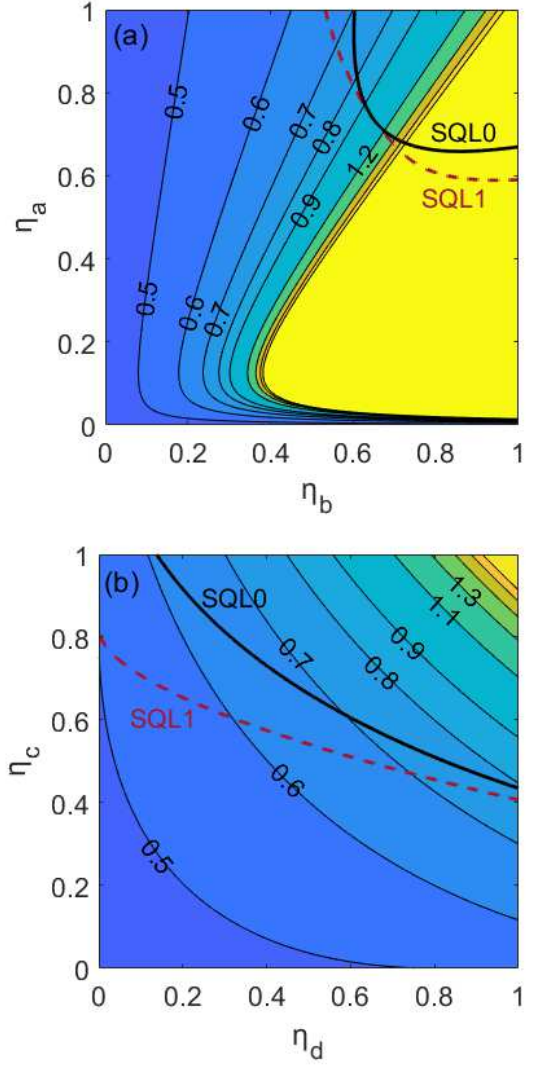}} \caption{ Results
of HD. SQL0 (solid line), SQL1 (dashed line) and the contour line of optimized
$G_{2}/G_{1}$ as a function of (a) $\eta_{a}$ and $\eta_{b}$ with $\eta
_{c}=\eta_{d}=1$, and (b) $\eta_{c}$ and $\eta_{d}$ with $\eta_{a}=\eta_{b}%
=1$. Parameters: $N_{c}=1000$, $T=R=0.5$, $G_{1}^{2}=5$. The phase
sensitivities in the area of upper right corner and within the line SQL0 or
SQL1 can beat the SQL.}%
\label{fig4}%
\end{figure}

Under the condition of lossless, the phase sensitivity is given by%
\begin{equation}
\Delta^{2}\phi^{HD}=\frac{A+B\cos\phi}{D\cos^{2}\phi},
\end{equation}
where%
\begin{align}
A &  =G_{2}^{2}g_{1}^{2}T^{2}+R^{2}G_{2}^{2}g_{1}^{2}+G_{1}^{2}g_{2}%
^{2}-2G_{1}G_{2}g_{1}g_{2}T+1/2,\nonumber\\
B &  =2G_{2}g_{1}R(G_{1}g_{2}-G_{2}g_{1}T)\text{, }D=2N_{c}G_{2}^{2}TR.
\end{align}
In general, due to $G_{1}g_{2}-G_{2}g_{1}T\geq0$, then $B\geq0$ is obtained.
Under the above conditions, whether the interferometer is balanced or not, we
can always get the optimal sensitivity at $\phi=\pi$. For balanced case
($G_{1}=G_{2}$), the optimal phase sensitivity $\Delta^{2}\phi^{HD}$ is
written as%
\begin{equation}
\Delta^{2}\phi_{opt}^{HD}=\frac{1}{4TRG^{2}N_{c}}.
\end{equation}
In Fig.~\ref{fig2}(a), the sensitivity $\Delta\phi_{opt}^{HD}$ can beat the
SQL. It can approach but cannot reach the QCRB. For the unbalanced case
($G_{1}\neq G_{2}$), the optimal $\Delta^{2}\phi^{HD}$ is written as
\begin{equation}
\Delta^{2}\phi_{opt}^{HD}=\frac{2\left(  G_{2}g_{1}-G_{1}g_{2}\right)  ^{2}%
+1}{4N_{c}G_{2}^{2}TR}.
\end{equation}
When the gain ratio $G_{2}/G_{1}\geq2$ and increases, the sensitivity
$\Delta\phi_{opt}^{HD}$ always reach the QCRB, as shown in Fig.~\ref{fig2}(b)
and Fig.~\ref{fig3}. For a given $G_{1}$, to obtain the optimal sensitivity
the value of $G_{2}$ should satisfy the following conditions%
\begin{equation}
\left(  \frac{g_{2}}{G_{2}}\right)  _{opt}=\frac{2G_{1}g_{1}}{2G_{1}^{2}-1}.
\end{equation}

In the presence of losses, from Eq. (\ref{error}) and Eq. (\ref{variance}) we
can get the optimal $\Delta\phi_{opt}^{HD}$ when $\phi=\pi$. On balanced case,
the phase sensitivity $\Delta\phi_{opt}^{HD}$ as a function of transimission
rates\ of two arms can be obtained, where the SQL denoted as SQL0 in the
contour figure. After optimizing the $G_{2}$, the phase sensitivity
$(\Delta\phi_{opt}^{HD})^{\prime}$ as a function of transimission rates\ of
two arms can also be obtained, where the SQL denoted as SQL1 in the contour
figure. For given $G_{1}$ the value of SQL0 and SQL1 is equal, the position of
SQL in the contour figure of the sensitivity versus transimission rates has changed.

When only considering the external loss $\eta_{a}$ and $\eta_{b}$ ($\eta
_{c}=\eta_{d}=1$) or internal loss $\eta_{c}$ and $\eta_{d}$ ($\eta_{a}%
=\eta_{b}=1$), the SQL0 (solid line), SQL1 (dashed line) and the contour line
of optimized $G_{2}/G_{1}$ as a function of $\eta_{a}$ and $\eta_{b}$ or
$\eta_{c}$ and $\eta_{d}$ are shown in Fig. \ref{fig4}(a) and Fig.
\ref{fig4}(b), respectively. The phase sensitivities in the area of upper
right corner and within the SQL lines in Fig. \ref{fig4}(a) and Fig.
\ref{fig4}(b) can beat the SQL. From these two figures, it is demonstrated
that for a given $G_{1}$ and transimission rates of two arms by optimizing
$G_{2}/G_{1}$, the small area between the SQL0 and SQL1 can still beat the SQL.

By Comparison, in Fig. \ref{fig4}(b) when only considering the internal loss
$\eta_{c}$ and $\eta_{d}$, the added area that can beat SQL between the SQL0
and SQL1 is larger. We obtain that the phase sensitivity can tolerate the
internal loss $\eta_{c}$ and $\eta_{d}$, and it can still beat SQL with about
$20\%$ of the photon loss of $\eta_{c}$ even if $\eta_{d}=0$. The reason is
that the internal loss only affects the output field of MZI. However, the
external loss affects not only the output field of MZI, but also the quantum
field of another combined beam.

\subsection{Intensity detection}

As shown in Fig~\ref{fig1}, we use the $\hat{n}_{32}=\hat{a}_{2}^{\dagger}%
\hat{a}_{2}+\hat{b}_{3}^{\dagger}\hat{b}_{3}$ as the detection variable. We
analyze the phase sensitivity at $\theta_{1}=0$, and $\theta_{2}=\pi$. The
slope is
\begin{align}
\frac{\partial\langle\hat{n}_{32}\rangle}{\partial\phi} &  =[2TR\eta_{b}%
\sqrt{\eta_{d}\eta_{c}}\left(  G_{2}^{2}+g_{2}^{2}\right)  \left(  N_{c}%
-g_{1}^{2}\right)  \nonumber\\
&  +4RG_{2}G_{1}g_{2}g_{1}\sqrt{\eta_{a}\eta_{b}\eta_{c}}]\sin\phi,
\end{align}
the fluctuation of the quadrature $\Delta\hat{n}_{32}$ is%
\begin{align}
\langle\Delta^{2}\hat{n}_{32}\rangle &  =N_{C}[\left\vert H_{3}H_{2}^{\ast
}\right\vert ^{2}+\left\vert H_{3}H_{3}^{\ast}\right\vert ^{2}+\left\vert
H_{3}H_{5}^{\ast}\right\vert +\left\vert H_{3}H_{6}^{\ast}\right\vert
^{2}\nonumber\\
&  +\left\vert H_{3}H_{7}^{\ast}\right\vert ^{2}+\left\vert H_{1}H_{3}^{\ast
}\right\vert ^{2}+\left\vert H_{4}H_{3}^{\ast}\right\vert ^{2}]+\left\vert
H_{1}H_{2}^{\ast}\right\vert ^{2}\nonumber\\
&  +\left\vert H_{1}H_{3}^{\ast}\right\vert ^{2}+\left\vert H_{1}H_{5}^{\ast
}\right\vert ^{2}+\left\vert H_{1}H_{6}^{\ast}\right\vert ^{2}+\left\vert
H_{1}H_{7}^{\ast}\right\vert ^{2}\nonumber\\
&  +\left\vert H_{4}H_{2}^{\ast}\right\vert ^{2}+\left\vert H_{4}H_{3}^{\ast
}\right\vert ^{2}+\left\vert H_{4}H_{5}^{\ast}\right\vert ^{2}+\left\vert
H_{4}H_{6}^{\ast}\right\vert ^{2}\nonumber\\
&  +\left\vert H_{4}H_{7}^{\ast}\right\vert ^{2},
\end{align}
where $H_{i}H_{j}^{\ast}=T_{i}T_{j}^{\ast}+M_{i}M_{j}^{\ast}$
$(i,j=1l,2l,3l,4,5,6,7).$

Under the condition of lossless, the phase sensitivity is given by%
\begin{equation}
\Delta\phi^{ID}=\left[  \frac{K}{I}+\frac{J^{2}+L\left(  \cos\phi+1\right)
^{2}}{I\sin^{2}\phi}\right]  ^{1/2},
\end{equation}
where subscript $ID$ indicates the intensity detection and%
\begin{align}
I  &  =\left[  2TR\left(  G_{2}^{2}+g_{2}^{2}\right)  \left(  N_{c}-g_{1}%
^{2}\right)  +4G_{1}G_{2}g_{1}g_{2}R\right]  ^{2},\nonumber\\
J  &  =G_{1}g_{1}\left(  G_{2}^{2}+g_{2}^{2}\right)  \left[  2TR-2+2TR\cos
\left(  \phi\right)  \right] \nonumber\\
&  +2G_{2}g_{2}\left(  G_{1}^{2}+g_{1}^{2}\right)  \left(  T-R\cos\phi
_{1}\right)  ,\nonumber\\
K  &  =\left[  2G_{2}G_{1}g_{2}-g_{1}\left(  G_{2}^{2}+g_{2}^{2}\right)
\right]  ^{2}TR\left(  N_{c}+1\right) \nonumber\\
&  +\left[  \left(  G_{2}^{2}+g_{2}^{2}\right)  G_{1}-2G_{2}g_{2}g_{1}\right]
^{2}TRN_{c}+4G_{2}^{2}g_{2}^{2}R^{2},\nonumber\\
L  &  =\left[  2G_{2}G_{1}g_{2}-g_{1}\left(  G_{2}^{2}+g_{2}^{2}\right)
\left(  T-R\right)  \right]  ^{2}TR\left(  N_{c}+1\right) \nonumber\\
&  +\left[  \left(  G_{2}^{2}+g_{2}^{2}\right)  G_{1}\left(  T-R\right)
-2G_{2}g_{1}g_{2}\right]  ^{2}TRN_{c}\nonumber\\
&  +4T^{2}R^{2}\left(  G_{2}^{2}+g_{2}^{2}\right)  ^{2}N_{c}.
\end{align}

\begin{figure}[ptb]
\includegraphics[scale=0.65,angle=0]{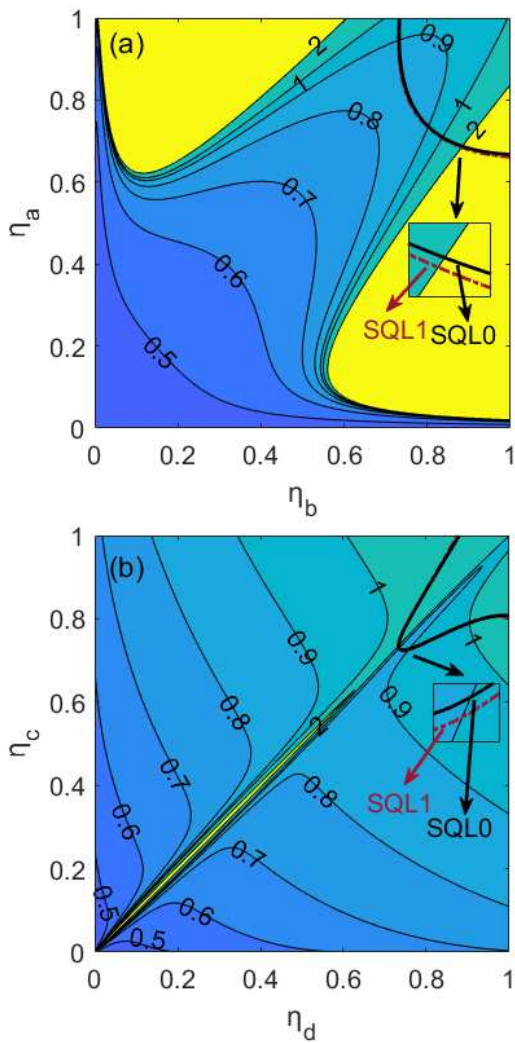} \caption{ Results of ID. SQL0
(solid line), SQL1 (dashed line) and the contour line of optimized
$G_{2}/G_{1}$ as a function of (a) $\eta_{a}$ and $\eta_{b}$ with $\eta
_{c}=\eta_{d}=1$, and (b) $\eta_{c}$ and $\eta_{d}$ with $\eta_{a}=\eta_{b}%
=1$. Parameters: $N_{c}=1000$, $T=R=0.5$, $G_{1}^{2}=5$. The phase
sensitivities in the area of upper right corner and within the line SQL0 or
SQL1 can beat the SQL.}%
\label{fig5}%
\end{figure}

Next, we analyze the results of balanced ($G_{1}=G_{2}$) and unbalanced cases
($G_{1}\neq G_{2}$). For balanced case, when $\phi\approx\pi$, we can simply
come to the following conclusion $J=0$, and $\frac{\left(  \cos\phi+1\right)
^{2}}{\sin^{2}\phi}\approx0$. Therefore the optimal sensitivity of $\Delta
^{2}\phi^{ID}$ is%
\begin{equation}
\Delta^{2}\phi_{opt}^{ID}\approx\frac{K}{I}\gtrsim\frac{1}{4TR\left(
G^{2}+g^{2}\right)  N_{c}}.
\end{equation}
Compared with the bound obtained from $F_{ave}$, the optimal sensitivity using
the ID can approach the corresponding QCRB, as the solid line shown in Fig.
\ref{fig2}(a). By comparing $\Delta\phi_{opt}^{HD}$ and $\Delta\phi_{opt}%
^{ID}$, we can obtain that the sensitivity of ID is superior than that of HD
in the balanced case, i.e. $\Delta\phi_{opt}^{ID}<\Delta\phi_{opt}^{HD}$. The
sensitivities of two approaches can beat the SQL and the sensitivity of ID can
reach the the QCRB.

For unbalanced case, when $\phi$ tends to $\pi$, the sensitivity of ID will
diverge and the optimal phase point is very near $\pi$ and dependent on the
values of $G_{1}$ and $G_{2}$, as shown in Fig. 2(b). The optimal value
$\phi_{opt}$ as a function of $G_{2}/G_{1}$ for ID is shown in Fig.~\ref{fig3}%
. For lossless case, when $G_{1}=G_{2}$ the detection results of ID is the
best. When the gain ratio $G_{2}/G_{1}$ is greater than $1$, the phase
sensitivity $\Delta\phi^{ID}$ always approaches but does not reach the QCRB.

Next, we research the effect of loss on the phase sensitivity $\Delta\phi
^{ID}$ and corresponding optimal gain ratio $(G_{2}/G_{1})_{opt}$. When only
considering the external loss $\eta_{a}$ and $\eta_{b}$ ($\eta_{c}=\eta_{d}%
=1$) or internal loss $\eta_{c}$ and $\eta_{d}$ ($\eta_{a}=\eta_{b}=1$), the
SQL0 (solid line), SQL1 (dashed line) and the contour line of optimized
$G_{2}$ as a function of $\eta_{a}$ and $\eta_{b}$ or $\eta_{c}$\ and
$\eta_{d}$ are shown in Fig. \ref{fig5}(a) and Fig. \ref{fig5}(b),
respectively. Different from HD method, by optimizing $G_{2}$ the ID method
can only increase a little area that can beat SQL. Comparing the two detection
methods, it is found that the method HD is more tolerant of photon losses.

\section{Conclusion}

In conclusion, we theoretically have derived the QCRB of ACMZI according to
the QFI obtained by the phase-averaging method. To approach the QCRB, we have
calculated the phase sensitivities with the method of ID and HD. Under the
condition of lossless, the phase sensitivity with the method of ID can
approach the QCRB in balanced case. In the presence of losses, the phase
sensitivity is reduced and associated optimal condition is also changed.
Within a certain loss range, we can adjust the gain parameters of the beam
recombination process to reduce the reduction in sensitivity and realize the
sensitivity can continue to beat SQL in an unbalanced situation. In the
presence of loss, comparing the results obtained by two different detection
methods, we found that the HD is more tolerant of internal loss.

\section{ACKNOWLEDGMENTS}

This work is supported by the National Key Research and Development Program of
China (2016YFA0302001); National Natural Science Foundation of China
(11974111, 11874152, 11654005); Shanghai Municipal Science and Technology
Major Project (Grant No. 2019SHZDZX01); the Shanghai talent program;
Fundamental Research Funds for the Central Universities.

\end{document}